\documentclass[preprint,11pt]{elsarticle}
\usepackage{multirow}
\usepackage[suffix=]{epstopdf}
\usepackage{float}
\usepackage[numbers]{natbib}
\usepackage{subfig}

\usepackage{array, makecell}
\usepackage{listings}
\usepackage{enumitem}
\usepackage[numbered,framed]{matlab-prettifier}
\usepackage{textcomp}
\usepackage{geometry}
\usepackage{mathtools}
\usepackage{color}

\usepackage{amssymb,amsmath}
\usepackage{amsfonts}
\usepackage{graphics}
\usepackage{amsthm}

\newcommand{\A}{\mathcal{A}}
\newcommand{\B}{\mathcal{B}}
\newcommand{\C}{\mathcal{C}}

\definecolor{mygreen}{RGB}{28,172,0} % color values Red, Green, Blue
\definecolor{mylilas}{RGB}{170,55,241}

\newtheorem{theorem}{Theorem}

\newtheorem{definition}[theorem]{Definition}

\begin{document}
\begin{frontmatter}
\title{Hidden and self-exited attractors in a heterogeneous Cournot oligopoly model}


%\titlerunning{Chaos control and anti-control of the heterogeneous Cournot oligopoly model}        % if too long for running head
\author[rm1,rm2]{Marius-F. Danca\corref{cor1}}
\cortext[cor1]{Corresponding author}
%\fntext[fn1]{This is the first author footnote.}
\author[rm3,rm4]{Marek Lampart}
\address[rm1]{Dept. of Mathematics and Computer Science, Avram Iancu University of Cluj-Napoca, Romania}
\address[rm2]{Romanian Institute of Science and Technology, 400487 Cluj-Napoca, Romania}
\address[rm3] {IT4Innovations, VSB~-~Technical University of Ostrava, 17.~listopadu 2172/15, 708 33 Ostrava, Czech Republic}
\address[rm4]{Department~of~Applied~Mathematics, VSB~-~Technical University of Ostrava,  17.~listopadu 2172/15, 708 33 Ostrava, Czech Republic}
%\email{marek.lampart@vsb.cz}
%ORCID: 0000-0001-6349-8553
\begin{abstract}In this paper it is numerically proved that a heterogeneous Cournot oligopoly model presents hidden and self-excited attractors. The system has a single equilibrium and a line of equilibria. The bifurcation diagrams show that the system admits several attractors coexistence windows, where the hidden attractors can be found. Intensive numerical tests have been done.
\end{abstract}
\end{frontmatter}

%%%%%%%%%%%%%%%%%%%%%%
\section{Introduction}\label{sec:1}
Since 1838, when A.~Cournot \cite{C} proposed the first treatment of oligopoly (a duopoly case), the theory of a market form in which a market has a dominant influence on a small number of sellers (oligopolists) was deeply researched.
The first and crucial additions to the theory were made by H.~von~Stackelberg \cite{S} and later on significances to the theory were done from a different point of view.

\medskip
As the Cournot-Nesh equilibria, of the corresponding game, reflects given oligopoly behaviour, its stability has to be investigated depending on the number of players and also on the way of the game modelling.
For the second case, it was given by \cite{T} (see also \cite{Pa} page 237) that the oligopoly model constructed under constant marginal costs with a linear demand function
is neutrally stable for three competitors and unstable for more than three competitors (for more details see \cite{P4}).
It is noted in \cite{P4} that linear demand functions are very easy to use, but they do not avoid negative supplies and prices,
so it is possible to use them only for the study of local behaviour.
Hence, nonlinear demand functions such as piecewise linear functions or other more complex functions were applied.
For duopoly by \cite{P}, and later by \cite{P2} for a triopoly using iso-elastic demand functions.
These types of demand function were later studied by \cite{A} and \cite{AA} for a nonlinear (iso-elastic) demand function and constant marginal costs
and it was concluded that this Cournot model for $n$ competitors is neutrally stable if $n=4$ and is unstable if the number of competitors is greater than five (see also \cite{P4}).
Finally, a complete characterization of the Cournot-Nesh stability was done in \cite{Lam1} depending on the number of competitors.

\medskip
All the above-given approaches were done for a homogeneous approach.
In heterogeneous decision mechanism, introduced in \cite{BN}, two different types of quantity setting players characterized by different decision mechanisms that coexist and operate simultaneously are considered.
In this case, competitors adaptively get used their choices towards the direction increasing their profits. This model's Cournot-Nesh equilibria stability was described showing is periodic and also chaotic regimes.
Moreover in \cite{BN2}, an addition to the foregoing approach was done where the role of the intensity of scenario choice was taken into consideration.

On the other side, hidden attractors represent an important recently introduced notion in applications because they might allow unexpected and potentially disastrous systems responses to some perturbations in a structure like a bridge or aircraft wing. However, except some examples of theoretical models (see e.g. \cite{opt,noua,zece}), there are no important investigations on hidden attractors in real and applied examples of chaotic maps.

The generally accepted definition which gives an attractors classification is
\begin{definition} \cite{unu,leo} \label{deff1}An attractor is called self-excited attractor if its basin of attraction intersects with any open neighborhood of an equilibrium, otherwise it is called hidden attractor.
\end{definition}

Sudden appearance of some hidden chaotic attractor, could represents a major disadvantage for the underlying system. Thus, the consequences could be dramatic such as in the case of pilot-induced oscillations that entailed the YF-22 crash in April 1992 and Gripen crash in August 1993 \cite{unspe}. It is understandable that identifying unwanted hidden chaotic behavior is a desirable phenomena. There exists the risk of the sudden jump from a desirable attractor to possible undesired behavior of some hidden attractor.
Recently, it has been shown that multistability is connected with the occurrence of hidden attractors.
If there are unstable fixed points, the basins of attraction of the hidden attractors do not touch them, being located far away from such points.
Note that if the system exhibits a chaotic or regular behavior while systems equilibria are stable, then the chaotic or regular underlying attractors are implicitly hidden.
Therefore, the stability of equilibria is important

For a hidden attractor, its attraction basins are not connected  with  unstable  equilibria.
Hidden attractors can be found in e.g. systems with no-equilibria or with stable equilibria \cite{unspe}.

 Also, as in the case of the studied discrete-time system in this paper, systems with infinite number of equilibria (also called line of equilibria), can admit hidden attractors. Systems with a line of equilibria are very few (see e.g. \cite{doispe,treispe}). Hidden attractors into an impulsive discrete dynamical system have been found in \cite{dd1}, where the case of a supply and demand economical system is studied.

The paper is organized as follows: Section 2 presents the considered oligopoly model, underlining equilibria stability, necessary in the study of hidden attractors, Section 3 deals with hidden and self-exited attractors, while Conclusion ends the paper.

\section{The heterogeneous Cournot oligopoly model}\label{sec:2}
%%%%%%%%%%%%%%%%%%%%%%

Consider the {\it heterogeneous Cournot oligopoly model} (HCOM) introduced in \cite{BN2} defined for identical quantity setting agents $\mathcal{N}=\{1, 2, \ldots ,N\}$ that compete in the same market for an homogeneous good, whose demand is summarized
by a linear inverse-demand function, or price function $P(Q)={\rm \max}\{a-bQ, 0\}$ ($P$ treats price as a function of quantity demanded). Denote by $q_{i}^n$ the quantity of goods that is generic $i$-th agent, with $i \in \mathcal{N},$ sells in the market
at time-period $n$. All the agents bear the same constant marginal production cost $c$, so that the generic $i$-th agent earns the profit
\begin{eqnarray*}\label{profit}
\pi_{i}=P(Q)q_i-cq_i.
\end{eqnarray*}

The oligopoly in this case, is characterized by introducing heterogeneous decision mechanisms, used to decide how much quantity of goods to produce, by considering a population structured into
two groups of agents of different kinds. The first group whose representative is denoted by $q_1$
 includes boundedly rational players that use {\it gradient} rule, that is called {\it gradient} player.
The second group with marked representant $q_2$ includes agents that adopt an imitation-based decision mechanism called {\it imitator} players.

The collective behavior of the whole heterogeneous population of $N$ players is described by the following $2$-dimensional non-linear autonomous discrete dynamical system \cite{BN2}:

\begin{equation}\label{HCOM}
\mbox{HCOM}:\left\{
\begin{array}{ccl}q_{1}^{n+1}&=&q_1^{n}+\gamma q_1^n(a-b((N(1-\omega)+1)q_1^n+\omega Nq_2^n)-c),\\
q_{2}^{n+1}&=&\displaystyle\frac{\pi_2^n}{\pi_2^n+\pi_1^n} q_2^n+\displaystyle\frac{\pi_1^n}{\pi_2^n+\pi_1^n} q_1^n,
\end{array}
\right.
\end{equation}
where
\begin{eqnarray}
\pi_1^n&=&(a-c-bN((1-\omega)q_1^n+\omega q_2^n))q_1^n,\\
\pi_2^n&=&(a-c-bN((1-\omega)q_1^n+\omega q_2^n))q_2^n,
\end{eqnarray}
%the system's parameters being set as in \cite{BaiNai2018} (Table \ref{tab:1}).

%Parameters $\gamma$ and also $\omega$ are considered as bifurcation parameters.

The system admits a line of equilibria
\[
X_0^*(0,q_2),~ q_2\geq 0,
\]
along the positive axis $(Oq_2)$, and also the single equilibrium

\[
X_1^*\Big(\frac{a-c}{b(N+1)},\frac{a-c}{b(N+1)}\Big).
\]

The stability of equilibria is stated by the following result

\begin{theorem}\cite{BN2}\label{tt}
\begin{itemize}
\item[i)]
Equilibrium $X_*^1$ is asymptotically stable if
\[
\omega \in\big(\Omega_1,\Omega_2\big),
\]
with
\[
\Omega_1=\frac{3}{2}\frac{N+1}{N}\Big(\frac{1}{2}-\frac{1}{\gamma(a-c)}\Big),~
\Omega_2=\frac{1}{2}\frac{N+1}{N}\Big(\frac{1}{\gamma(a-c)}+1\Big).
\]
\item[ii)]
Equilibria $X_0^*(0,q_2)$ are stable for $q_2>q_2^*=\frac{a-c}{b\omega N}$.
\end{itemize}
\end{theorem}

Note that the stability, established by Theorem \ref{tt} is locally\footnote{In \cite{BN2} this relation seems to be wrong.}.
\bigskip

\noindent \emph{\underline{Graphical interpretation}}
\vspace{3mm}

Consider in the parameter space $(\gamma,\omega)$, the latice domain $\mathcal{D}=[0.35,0.525]\times[0.329,0.721]$ with corners $M_1(0.35,0.329),M_2'(0.525,0.329),M_3'(0.525,0.791),M_4(0.35,0.791)$ (Fig. \ref{fig1}).

All numerical tests in this paper have been done for the constants $a=10$, $b=1$, $c=1$, $N=5$ and the parameters $\gamma$, $\omega$ as bifurcation parameters. For all these values, the equilibrium $X_1^*=(1.5,1.5)$.

As functions of $\gamma$, $\Omega_1$, with graph $\Gamma_1$ and $\Omega_2$, with graph $\Gamma_2$, are bifurcation curves representing the flip bifurcation and Neimark-Saker bifurcation, respectively, of $X_1^*$. For all $(q_1,q_2^*)$, $X_0^*$ suffers a Neimark-Saker bifurcation \cite{BN2}.

The stability of points $X_{0,1}^*$ reads as follows: the dark green area $S\subset\mathcal{D}$ limited by the curves $\Gamma_{1,2}$ and lines $\gamma=0.35$ and $\gamma=0.525$ and defined by the corners $M_1(0.35,0.329),\\M_2(0.525,0.519),M_3(0.525,0.727),M_4(0.35,0.791)$ (Fig. \ref{fig1}), contains the parameter sets $(\gamma,\omega)$ which generate the stability of the equilibrium $X_1^*$, while the light green areas (outside the curves $\Gamma_{1,2}$) represent the instability sets of the equilibrium $X_1^*$.

The vertical dotted line through $\gamma=0.48$ and the horizontal dotted line through $\omega=0.4$ represent the bifurcation diagrams with respect $\omega$ and $\gamma$, respectively.

\section{Attractors coexistence and hidden attractors}

As know, due to the uniqueness of the solutions of the Initial Value Problem (IVP) \eqref{HCOM} (ensured by the explicit form of the system equations), for a fixed parameter value, to each initial condition corresponds uniquely an attractor. In this paper one consider \emph{numerical attractors}, obtained by numerical integration of the IVP, after sufficiently large transients removed.
For simplicity, hereafter, by attractor one understands the underlying numerical attractor.
Therefore, in a bifurcation diagram ($BD$) generated with fixed initial condition, to every bifurcation parameter value corresponds a unique attractor represented in the $BD$ as a vertical line, composed from a set of isolated points (periodic attractors), or a band of infinity of points (quasiperiodic or chaotic attractors)\footnote{To be focused, in this paper every chaotic orbit is as usual understood
as an orbit approaching a chaotic attractor, even if the exiting (interior
and exterior) crises might imply chaotic but non-attracting sets (see e.g. the non-attracting chaotic set after the saddle-node bifurcation of the logistic map for $r\approx4.83$).}.

Because in this paper every $BD$ is generated with two different initial conditions ($IC$), namely $IC_1=(1,1.5)$ and $IC_2=(1.75,1.5)$, one obtain two sets of attractors represented by vertical lines of points in the $BD$ denoted with fraktur letters with index $1$ and $2$: $(\mathfrak{A}_1, \mathfrak{A}_2)$, $(\mathfrak{B}_1, \mathfrak{B}_2)$ and so on. Therefore, at the considered resolution of $800$ points on the bifurcation parameter axis, correspond two sets of $800$ attractors, denoted with calligraphic letters indexed $1$ or $2$ depending the initial conditions $IC_{1,2}$, $\A_1\in \mathfrak{A}_1$, $\B_1\in\mathfrak{B}_1$, $\A_2\in\mathfrak{A}_2$, $\B_2\in\mathfrak{B}_2$ and so on. All these attractors are function of the considered bifurcation parameter, $\A_1=\A_1(p)$, $\B_1=\B_1(p)$ and so on, the parameter $p$ being either $\gamma$ or $\omega$, and are plotted red and blue corresponding to $IC_1$, or to $IC_2$, respectively.

For simplicity, hereafter one drops the parameter $p$ in attractors notation and all attractors in this paper are generated by starting from one of the initial conditions $IC_1$, or $IC_2$. Note that every attractor can also be generated from the indicated initial conditions from underlying attraction basins, denoted $q_0$.

For some parameter ranges, the two sets of attractors could be different (inside the coexistence windows), when the existence of hidden attractors is possible, or identic (outside coexistence windows).

Within the studied coexistence windows, the system \eqref{HCOM} presents three different kind of attractors: periodic attractors or limit cycles, quasiperiodic attractors and chaotic attractors.

The tools utilized in this paper to identify attractors are: BDs, time series, planar phase representations, the maximal local finite-time Lyapunov exponent $\lambda$, the output $K$ of the 0-1 test for chaos (see e.g. \cite{01}) and Power Spectrum Density (PSD). Because the PSD is two-sided symmetric, only the left-side is considered. The numerical integration of the system \eqref{HCOM} has been effectuated for $n=3000$ iterations.

Following Definition \ref{deff1}, the algorithm used to detect numerically hidden and self-excited attractors of the considered system \eqref {HCOM} is presented in the diagram in Fig \ref{fig2}.
In systems in spaces with higher dimension with unstable equilibria, the attraction basins are chosen usually as planar sections containing unstable equilibria. The case of three-dimensional neighborhoods of the Fabrikant-Rabinovich system is treated in \cite{dd2}. As the diagram shows, the main steps in finding hidden attractors bases on testing if the analyzed attractor has initial conditions within no matter how small neighborhoods of all unstable equilibria ($X_0^*$, $X_1^*$). If there exists a neighborhood of at least one of the unstable equilibria containing initial conditions of the considered attractor, the attractor is self-excited. Otherwise, if the attraction basin does not intersect any of unstable equilibria, the attractor is hidden.

Because the case of stability of both equilibria is trivial (for example the case of chaotic attractors which, in this case, are all hidden by Definition \ref{deff1}), one considers the complicated case of unstable equilibria, when each equilibrium must be analyzed.

The main step in verifying if the attractors are hidden or self-excited, following the algorithm in Fig. \ref{fig2}, is to check neighborhoods of equilibria $X_{0,1}^*$. Precisely, one has to verify the connection of the attraction basins of the considered attractor with both equilibria. For this purpose one examine neighborhoods of both equilibria $X_0^*$ and $X_1^*$, considered separately for clarity
(see e.g. Figs. \ref{fig6}). The figures in Figs. \ref{fig6} represent latices of points $(q_1,q_2)$, considered as initial conditions for numerical integration of the IVP \eqref{HCOM}, containing equilibria $X_0^*$ and $X_1^*$, respectively. For $X_0^*$, which is a line of equilibria, the region containing the equilibrium, is a rectangular neighborhood with width $1e-5$ and height taken so that it includes $q_2^*$. Thus, the neighborhood contains a part of the line equilibria $X_0^*$ including the critical point $q_2^*$ (see Theorem \ref{tt} ii)).
Conform to Theorem \ref{tt} ii) the yellow points with $q_2>q_2^*$, represent initial conditions leading to the vertical axis, which is attractive, while for $q_2<q_2^*$, $X_0^*$ is repulsive. For $X_1^*$, the examined region is a squared lattice with side 1 centered on $X_1^*$. On both neighborhoods, red plot represents the initial conditions leading to attractors of the first set, corresponding to $IC_1$, while blue plot are the points leading to attractors belonging to the second coexisting set of attractors corresponding to $IC_2$.
Black points represent the divergence points, for which the system is unbounded.

\subsection{Hidden $\gamma$-attractors}
\begin{figure}
\begin{center}
\includegraphics[scale=0.43]{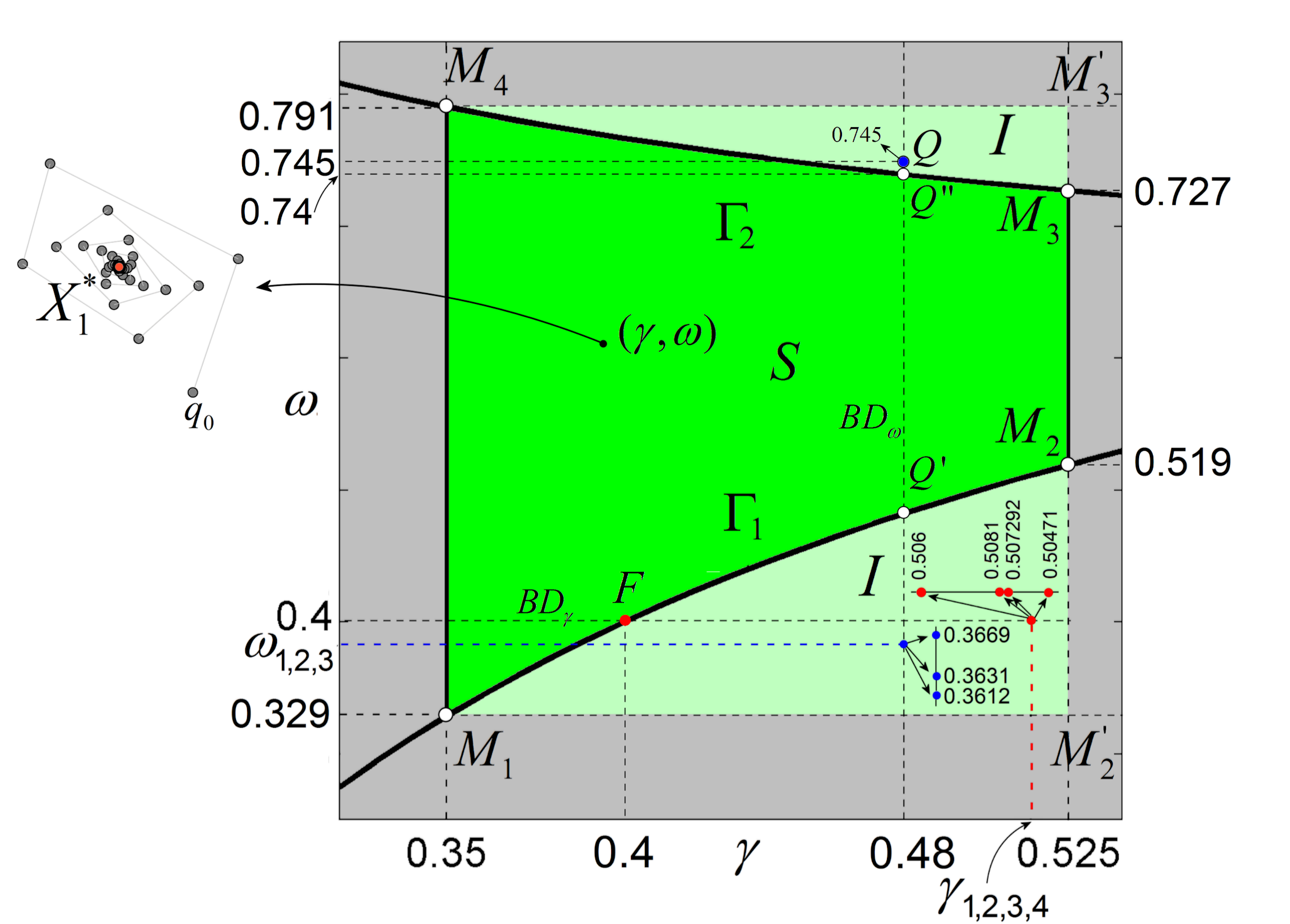}
\caption{The explored rectangular area $M_1M_2'M_3'M_4$ on the parameters plane $(\gamma,\omega)$. Dark green area $M_1,M_2,M_3,M_4$ represents the area where $X_1^*$ is stable. Parameters points $(\gamma,\omega)$ within this area generate counterclockwise orbits spiralling toward the stable equilibrium $X_1^*$. Light green represents the instability domain of $X_1^*$. Curves $\Gamma_{1,2}$ represent the limits stability. $\Gamma_1$ is the flip bifurcation curve and $\Gamma_2$ the NS bifurcation curve. Horizontal dotted line through $\omega=0.4$ represents the considered bifurcation line versus $\gamma$, $BD_\gamma$. The near red points represent the studied values of $\gamma$ in this bifurcation diagram. Vertical dotted line through $\gamma=0.48$ represents the bifurcation diagram versus $\omega$, $BD_\omega$. Blue near points on this line represent the values of $\omega$ studied in this bifurcation diagram. Points $F$, $Q'$, and $Q''$, $Q$ are points of flip and NS bifurcations, respectively.}
\label{fig1}
\end{center}
\end{figure}

\begin{figure}
\begin{center}
\includegraphics[scale=0.45]{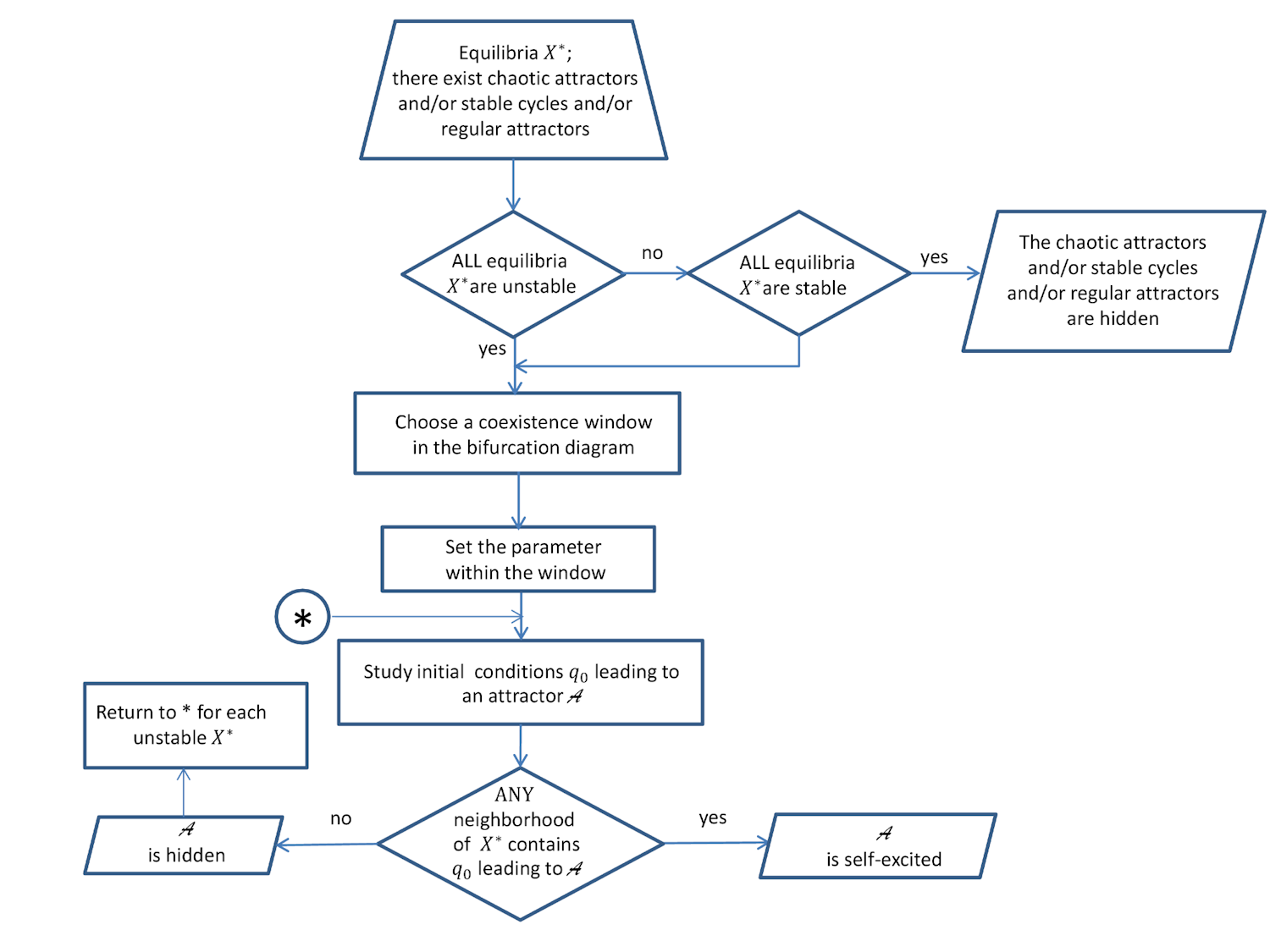}
\caption{Algorithm utilized in this paper to identify hidden attractors.}
\label{fig2}
\end{center}
\end{figure}

\begin{figure}
\begin{center}
\includegraphics[scale=0.35]{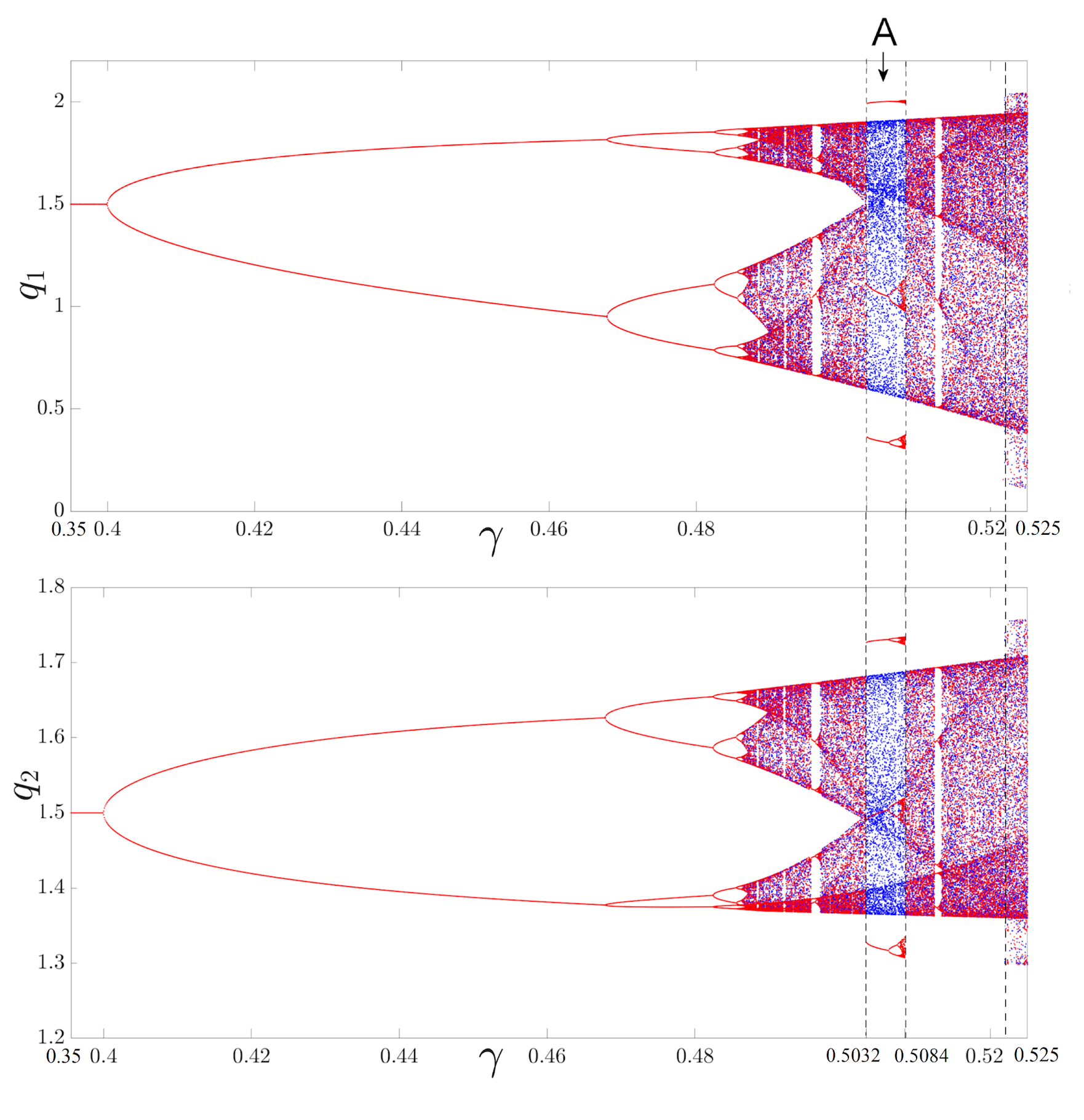}
\caption{Bifurcation diagram of the system \eqref{HCOM} versus $\gamma$, $BD_\gamma$, for both components $q_{1,2}$, for $\omega=0.4$. }
\label{fig3}
\end{center}
\end{figure}

Consider the $BD$ of $q_1$ and $q_2$ vs $\gamma\in[0.35,0.525]$ for $\omega=0.4$, denoted with $BD_\gamma$ (dotted line through $\omega=0.4$ in Fig. \ref{fig1}). Note that at $\gamma=0.4$, the diagram crosses the curve $\Gamma_1$ (point $F$ in Fig. \ref{fig1}), marking the first flip bifurcation of $X_1^*$, which at $\gamma=0.4$ looses his stability (see also Figs. \ref{fig3}).

Due to symmetry of the $BD$s with respect $\gamma$ and $\omega$ axis, for simplicity, hereafter only the component $q_1$ is considered.

\begin{figure}
\begin{center}
\includegraphics[scale=.8]{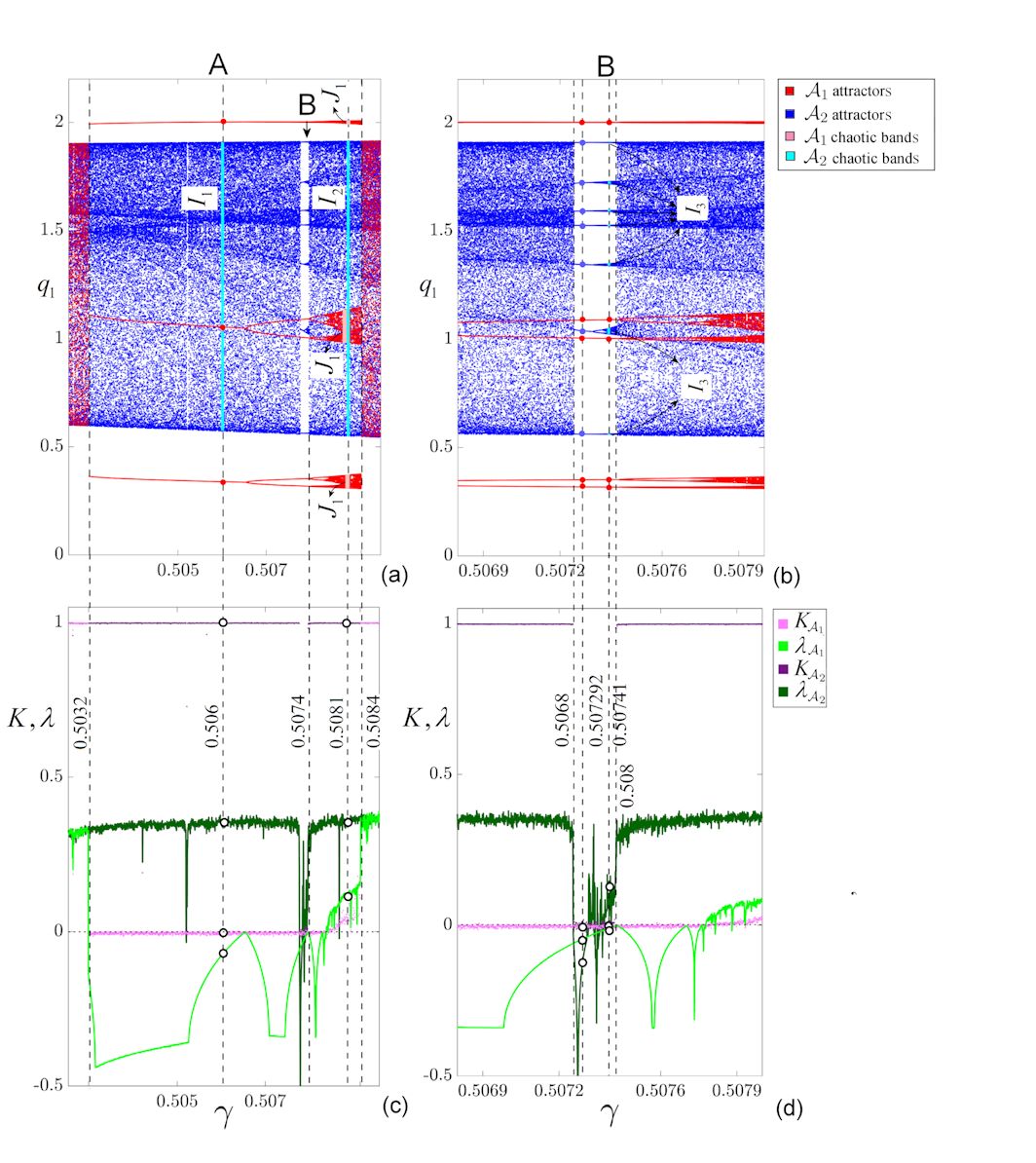}
\caption{Zoomed areas of the bifurcation diagram $BD_\gamma$. (a) Zoomed area for $\gamma\in [0.5032,0.5084]$; (b) Zoomed area for $\gamma\in [0.5068, 0.508]$; (c) $K$ and $\lambda$ for the zoomed area $A$; (d)  $K$ and $\lambda$ for the zoomed area $B$.}
\label{fig4}
\end{center}
\end{figure}

As specified bellow, attractors belonging to $\mathfrak{A}_{1}$ are denoted with $\mathcal{A}_{1}$, while the attractors of $\mathfrak{A}_{2}$ are denoted with $\A_2$ (red and blue plot, respectively, in Figs. \ref{fig4} (a), (b)).

Because the system dynamics related to hidden attractors regard mainly unstable equilibria, one consider on the $BD_\gamma$ the values $\gamma>0.4$, where the equilibrium $X_1^*$ is unstable (see Fig. \ref{fig1}) and where there are the coexistence window denoted $A$ and the zoomed window $B$, delimited by $\gamma\in (0.5032,0.5084)$ and $\gamma\in(0.507292, 0.50741)$, respectively (Figs. \ref{fig4} (a), (b)).
The window $A$ starts with an exterior crisis at $\gamma=0.5032$, after which a cascade of flip bifurcations leads to the chaotic behavior of the attractor $\mathcal{A}_1$, and ends with an interior crisis at $\gamma=0.5084$.

The window $B$ starts with an exterior crisis at $\gamma=0.5068$, where begins a cascade of flip bifurcations of both attractors $\mathcal{A}_{1,2}$, the bifurcations of $\mathcal{A}_2$ being ``delayed'' with respect $\gamma$, compared to the bifurcations of the attractor $\mathcal{A}_1$. At $\gamma=0.508$ the window ends with an interior crisis of the attractor $\mathcal{A}_2$.

Consider first $\gamma=\gamma_1=0.506$ (Fig. \ref{fig4} (a)) to which correspond two coexisting attractors, $\mathcal{A}_1\in\mathfrak{A}_1$ and $\mathcal{A}_2\in\mathfrak{A}_2$. The type of the attractor $\mathcal{A}_1$ is revealed by the three red bullet points in $BD_\gamma$ (Fig. \ref{fig4} (a)), $\lambda$, which is negative (bullet on the light green curve in Fig. \ref{fig4} (c)) and $K$ which is $0$ (bullet on the light magenta curve in Fig. \ref{fig4} (c)). Moreover, the time series (Fig. \ref{fig5} (a)) and the phase plot (Fig. \ref{fig5} (c)) indicate that the attractor $\mathcal{A}_1$ is period-3 (the three numbered red bullets in Fig. \ref{fig5} (a) and (c)).  The attractor $\mathcal{A}_2$ presents a so called \emph{chaotic band} (light blue band in Fig. \ref{fig4} (a)). $\lambda$ is positive (bullet on dark green curve in Fig. \ref{fig4} (c)) and $K\approx 1$ (bullet on the dark magenta curve in Fig. \ref{fig4} (c)). The time series (Fig. \ref{fig5} (b)) and the chaotic band appearing in the phase plot as projection of the attractor on $q_1$ axis (Fig. \ref{fig5} (c)) indicate the chaotic characteristic of $\mathcal{A}_2$.

For $\gamma=\gamma_2=0.5081$ (Fig. \ref{fig4} (a)), both attractors, $\mathcal{A}_1$ and $\A_2$, are chaotic, as shown by the $BD_\gamma$, $\lambda$, $K$ (Figs. \ref{fig4} (c)), the time series in Figs \ref{fig5} (d), (e) and the phase plots in Fig. \ref{fig5} (f). While the chaotic attractor $\mathcal{A}_1$ contains three chaotic bands (purple lines in Fig. \ref{fig4} (a) and Fig. \ref{fig5} (a), (f)), the chaotic attractor $\mathcal{A}_2$ is composed by a single chaotic band (light blue in Fig. \ref{fig4} (a) and Fig. \ref{fig5} (e) and Fig. \ref{fig5} (f)). Note that the three chaotic bands born from the previous stable three red points in Fig. \ref{fig5} (a), (c) which lost the stability.

At $\gamma=\gamma_3=0.507292$, value which can be viewed in Fig. \ref{fig4} (b), the two corresponding attractors $\mathcal{A}_1$, and $\mathcal{A}_2$, are two periodic attractors. The stable cycle $\mathcal{A}_1$ is a period-6 attractor (the apparent cycle point at about $q_1=2$, marked with * in the time series (Fig. \ref{fig5} (g)), is actually a superposition of two points of the cycle: points 2 and 5 in Fig. \ref{fig5} (i)). The other stable cycle, $\mathcal{A}_2$, is a period-7 attractor. The periods are revealed by the red and blue bullets, respectively, in the $BD_\gamma$, negative $\lambda$ and zero $K$ (Fig. \ref{fig4} (d)).

The last considered value, $\gamma=\gamma_4=0.50741$ generates the period-6 attractor $\mathcal{A}_1$ (the six red bullets in Fig. \ref{fig5} (j)) and the six pieces chaotic attractor $\mathcal{A}_2$, which presents six light blue chaotic bands (see Fig. \ref{fig5} (k) and the phase plot in Fig. \ref{fig5} (l)).
The periodic characteristic of $\mathcal{A}_1$ is underlined by the negative $\lambda$ (light green, Fig. \ref{fig4} (d)) and the zero value of $K$ (light magenta, Fig. \ref{fig4} (d)). Again, at $q_1=2$ there are two overplotted points (3 and 6 in Fig. \ref{fig5} (j) and (l)).

The analysis made in Figs. \ref{fig6} for $\gamma$-attractors, show that for every considered cases of $\gamma$, the attraction basins of attractors $\mathcal{A}_1$ for $\gamma\in\{\gamma_1,\gamma_2,\gamma_3,\gamma_4\}$ (red plot) have no connection with $X_1^*$ or $X_0^*$ and, therefore, they are hidden, while attractors $\mathcal{A}_2$, for $\gamma\in\{\gamma_1,\gamma_2,\gamma_3,\gamma_4\}$, have initial conditions which intersect any neighborhood of $X_1^*$ (blue plot), and, therefore, are self-excited.

Note that, for the considered $\gamma$ values, attractors $\A_1$ and $A\_2$ can be generated, beside $IC_1$ and $IC_2$, from indicated initial conditions $q_0$ too. Also, the critical value of $q_2^*$ is $q_2^*\approx 4.5$.

Summarizing:

\begin{enumerate}[noitemsep]
\item
 For $\gamma=\gamma_1$, the period-3 stable cycle $\mathcal{A}_1$ is hidden, while the chaotic attractor $\mathcal{A}_2$ is self-excited (Fig. \ref{fig6} (a));
\item For $\gamma=\gamma_2$, the chaotic attractor $\mathcal{A}_1$ is hidden, while the chaotic attractor $\mathcal{A}_2$ is self-excited (Fig. \ref{fig6} (b));
\item For $\gamma=\gamma_3$, the period-6 stable cycle $\mathcal{A}_1$ is hidden, while the stable period-7 stable cycle $\mathcal{A}_2$ is self-excited (Fig. \ref{fig6} (c));
\item For $\gamma=\gamma_4$, the period-6 stable cycle $\mathcal{A}_1$ is hidden, while the chaotic attractor $\mathcal{A}_2$ is self-excited (Fig. \ref{fig6} (d)).
\end{enumerate}

\subsection{Hidden $\omega$-attractors}

Consider the $BD$ versus $\omega$, $BD_\omega$, for $\omega\in[0.329,0.791)$ and $\gamma=0.48$ generated with the same initial conditions, $IC_1$ and $IC_2$ (Fig. \ref{fig7}). Like the $BD_\gamma$, the $BD_\omega$ crosses the stability and instability domains $S$ and $I$ (Fig. \ref{fig1}). Compared to the $BD_\gamma$ which intersects only the flip bifurcation curve, $\Gamma_1$, $BD_\omega$ intersects the NS bifurcation curve, $\Gamma_2$, at points $Q'$ and $Q''$, as well, the ingredient necessary to quasiperiodic oscillations. At $\omega=0.48325$, $X_1^*$ becomes stable (Fig. \ref{fig1}).

Note that for $\omega\in(0.48325,0.73855)$, range of $\omega$ which starts with the last reverse flip bifurcation (point $Q'$) and ends at the first NS bifurcation (Point $Q''$), the system dynamics do not depend on $\omega$, fact which could represents a useful system characteristic.
Denote the two coexisting sets of attractors with $\mathfrak{B}_1$ and $\mathfrak{B}_2$ with corresponding elements, attractors $\B_1$ and $\B_2$, respectively.

The considered coexistence windows starts with successive reversed flip bifurcations (period halving bifurcation) of the attractors $\mathcal{B}_1$, while the attractors $\mathcal{B}_2$ remain chaotic for a large parameter range $\omega\in (0.3608, 0.3681)$. The window ends with an interior crisis of $\mathcal{B}_1$.

The windows of interest generated by the parameter $\omega$ are denoted by $C$, his successive zoomed area $D$ and $E$ (Fig. \ref{fig7} (a), (b), (c)).

Within area $C$ and his zoomed area $D$, one consider three representative cases: $\omega_1=0.3612$, $\omega_2=0.3631$ and $\omega_3=0.3669$.

For $\omega=\omega_1$, the values of $\lambda$, and $K$ suggest chaotic dynamics for both underlying attractors, $\mathcal{B}_1$ and $\mathcal{B}_2$ (Fig. \ref{fig7} (c)). The time series in Fig. \ref{fig8} (a) and the phase plot (Fig. \ref{fig8} (b)) shows that the attractor $\mathcal{B}_1$ produces three chaotic bands $V_1$ (magenta plot). The attractor $\mathcal{B}_2$ presents a single large chaotic band (see Fig. \ref{fig8} (b) and phase plot in Fig. \ref{fig8} (c), light blue plot).

For $\omega=\omega_2$, by a reverse flip bifurcation, the former attractor $\mathcal{B}_1$, transforms into a stable period-6 cycle (see Fig. \ref{fig8} (d) and (f) where elements 1 and 4 have  the same $q_1$ value), while the attractor $\mathcal{B}_2$ remains chaotic but with a reduced size of the underlying chaotic band (Fig. \ref{fig8} (e), (f)).

At the last considered value $\omega=\omega_3$, by reverse flip bifurcation both attractors, $\mathcal{B}_1$ and the attractor $\B_2$, transform in stable cycles of period-3 and a period-14 cycle, respectively (Figs. \ref{fig8} (g), (h), (i)).

To see which $\omega$-attractors are hidden, one applies the algorithm presented in Fig. \ref{fig2}, for each considered case of $\omega$.

Again, for the considered $\omega$ values, attractors $\B_1$ and $\B_2$ can be generated, beside $IC_1$ and $IC_2$, from indicated initial conditions $q_0$ in the attraction basins too (Figs. \ref{fig9}). The critical value of $q_2^*$ is $q_2^*\approx 4.9$.

Attractors $\B_1$ are hidden (the red plot corresponding to the ICs of the attractor $\B_1$, indicates that the attraction basins do not touch equilibria), while $\B_2$ are self-excited.

\begin{enumerate}[noitemsep]
\item
For $\omega=\omega_1$, the chaotic attractor $\mathcal{B}_1$ is hidden, while the chaotic attractor $\mathcal{B}_2$ is self-excited;
\item
For $\omega=\omega_2$, the period-6 cycle $\mathcal{B}_1$ is hidden, while the chaotic attractor $\mathcal{B}_2$ is self-excited;
\item
For $\omega=\omega_3$, the period-3 cycle $\mathcal{B}_1$ is hidden, while the period-14 cycle $\mathcal{B}_2$ is self-excited;
\end{enumerate}

\emph{Quasiperiodic attractors}

An interesting particular case of $\omega$-attractors unlike the previous cases, is represented by the coexisting window $E$, defined by $\omega\in [0.7423,0.75216]$ (Fig. \ref{fig10} (a)).

Denote the two sets of attractors by $\mathfrak{C}_1$ and  $\mathfrak{C}_2$ with elements $\C_1$ and $C_2$, generated as for all cases from $q_0=(1.6,1.4)$.

Consider in this window $\omega=0.745$ (Fig. \ref{fig10} (a)). In this case $\lambda$ is negative for $\C_1$ and zero for $\C_2$, and $K$ is zero for both attractors (Fig. \ref{fig10} (b)) fact which indicate that $\C_2$ is quasiperiodic, while $\C_1$ period-4 cycle. This conclusion is sustained by the four red bullets in time series in Fig. \ref{fig11} (a), and the grey quasiperiodic band in Fig. \ref{fig11} (b) and also phase plot in Fig. \ref{fig11} (c). Note that the fact the points of the quasiperiodic curve in the phase plot tend to fill the entire closed quasiperiodic orbit (invariant circle), indicates that the orbit neither closes nor repeats itself. The quasiperiodicity ie revealed also by the PSD which shows clearly that the periodic orbit ($\C_1$) presents the first fundamental frequency $f_0$ and the harmonic $f_1$ situated at distance $\delta_1$ (Fig. \ref{fig11} (d)). Regarding $\C_2$, because of the NS bifurcation, which generally generates quasiperiodicity, a new set of subharmonics born, like $f_{01}$, close to first frequencies $f_0$ and $f_1$, at smaller distance distances $\delta_{01}$.

The attractor $\C_1$ is hidden, while the quasiperiodic orbit $\C_2$ is self-excited (see the attraction basins in Fig. \ref{fig12} (a)). Compared to previous cases, the attraction basins in this case, have more complicated shape and remember the riddled attraction basins (see e.g. \cite{rid}), when any arbitrary neighborhoods of every points of the attraction basin seems to contain points from some another basin (see e.g. circled regions of $X_{0,1}^*$ neighborhoods).

As can be seen, there exist thin yellow strips of points $(q_1,q_2)$ with $q_2<q_2^*$ for which $X_0^*$ is attractive, in contradiction with Theorem \ref{tt} ii). For example for the initial condition $q_0=(0.0005347,0.4380547)$, with $q_2=0.4380547<2.42=q_2^*$, the orbit is attracted by the vertical axis of equilibria $X_0^*$ (Fig. \ref{fig12} (b)). The second component of the orbit tends to a value situated beyond $q_2^*$, where $X_0^*$ is stable (see $q_2\approx3.5$ in Fig. \ref{fig12} (b)), but the initial condition belongs to instability domain established by Theorem \ref{tt} i). Also, all considered neighborhoods of $X_0^*$ reveal the fact that $q_2^*$ is not constant with respect $q_1$, but a function of $q_1$ too and, therefore, the graph of $q_2^*$ (separatrix between the yellow and blue domain) is not a constant horizontal line.
These apparent contradictions, could be related to the locally character results of the stability given by Theorem \ref{tt}.

All results are presented in Table \ref{tabb}.

\begin{table}
\small
$\begin{array}{lllll}
\multicolumn{1}{c}{\omega } & \multicolumn{1}{c}{\gamma} & \makecell{Hidden\\attractors}&\makecell{Self$-$excited\\attractors}&\multicolumn{1}{c}{Figures}\\\hline
%\omega &\gamma&Hidden~attractors & Self-excited~attractors & Figs \\ \hline
\multirow{4}{*}{$\omega =0.4$} &\multicolumn{1}{|l}{\gamma =0.506} & \mathcal{A}_1~period$-3$ &
\mathcal{A}_2~chaotic &Figs. ~5 ~(a)-(c)  \\
& \multicolumn{1}{|l}{\gamma =0.5081} & \mathcal{A}_1~chaotic & \
\mathcal{A}_2~chaotic~ & Figs.~ 5~ (d)-(f) \\
& \multicolumn{1}{|l}{\gamma =0.507292} & \mathcal{A}_1 ~period$-6$&
\mathcal{A}_2~period$-7$ & Figs.~ 5~ (g)-(i) \\
& \multicolumn{1}{|l}{\gamma =0.50741 }& \mathcal{A}_1 ~period$-6$~ & \mathcal{A}_2 ~chaotic& Figs.~5~ (j)-(l)\\\hline
\omega=0.3612& \multicolumn{1}{|l}{\multirow{4}{*}{$\gamma =0.48$}} & \mathcal{B}_1~ chaotic & \mathcal{B}_2 ~chaotic& Figs.~ 8~ (a)-(c)\\
\omega= 0.3631& \multicolumn{1}{|l}{} & \mathcal{B}_1 ~period$-6$ & \mathcal{B}_2 ~chaotic&Figs.~ 8~ (d)-(f)\\
\omega=  0.3669&\multicolumn{1}{|l}{}  & \mathcal{B}_1 ~period$-3$~ & \mathcal{B}_2 ~period$-14$&Figs.~ 8~ (g)-(i)\\
\omega=  0.745& \multicolumn{1}{|l}{} & \mathcal{C}_1 ~period$-4$~ & \mathcal{C}_2 ~quasiperiodic&Figs.~11~ (a)-(c)\\
&
\end{array}$
\caption{Hidden and self-excited attractors of the system \eqref{HCOM}.}.\label{tabb}
\end{table}

%\begin{table}
%\centering
%\caption{Parameters of IHCOM (\ref{eq:2}).}
%{\begin{tabular}{ccccc} %\toprule
%\hline
%a & b  &  c & N & $\omega$ \\ %\midrule
%\hline
%10& 1 &  1 & 5 & 0.4 \\
%$b_{\varphi}$ & 0.001 Nms rad$^{-1}$&  the outer damping coefficient of the impact body (rotational motion) \\
%$f$ & 0.2 Nms rad$^{-1}$&  coefficient of friction\\
%$a$ & 100 s m$^{-1}$& mathematical constant defining the shape of the friction characteristic\\ %\bottomrule
%\hline
%\end{tabular}}
%\label{tab:1}
%\end{table}
\vspace{3mm}
\textbf{Acknowledgements}
This work was supported by The Ministry of Education, Youth and Sports from the National Programme of Sustainability (NPU II) project ``IT4Innovations excellence in science -- LQ1602";
by The Ministry of Education, Youth and Sports from the Large Infrastructures for Research, Experimental Development and Innovations project ``IT4Innovations National Supercomputing Center -- LM2015070";
by SGC grant No. SP2020/137 ``Dynamic system theory and its application in engineering", VSB - Technical University of Ostrava, Czech Republic,
Grant of SGS No. SP2020/114, VSB - Technical University of Ostrava, Czech Republic.

\section{Conclusions}

In this paper hidden and self-exited attractors of a discrete heterogeneous Cournot oligopoly model were numerically found. The system proved to have extremely rich dynamics including attractors coexistence. All studied coexistence windows embed hidden attractors and self-excited attractors. To identify hidden attractors, one analyze the neighborhoods of unstable equilibria in order to see they have connections with the considered attractors.

\begin{figure}
\begin{center}
\includegraphics[scale=.45]{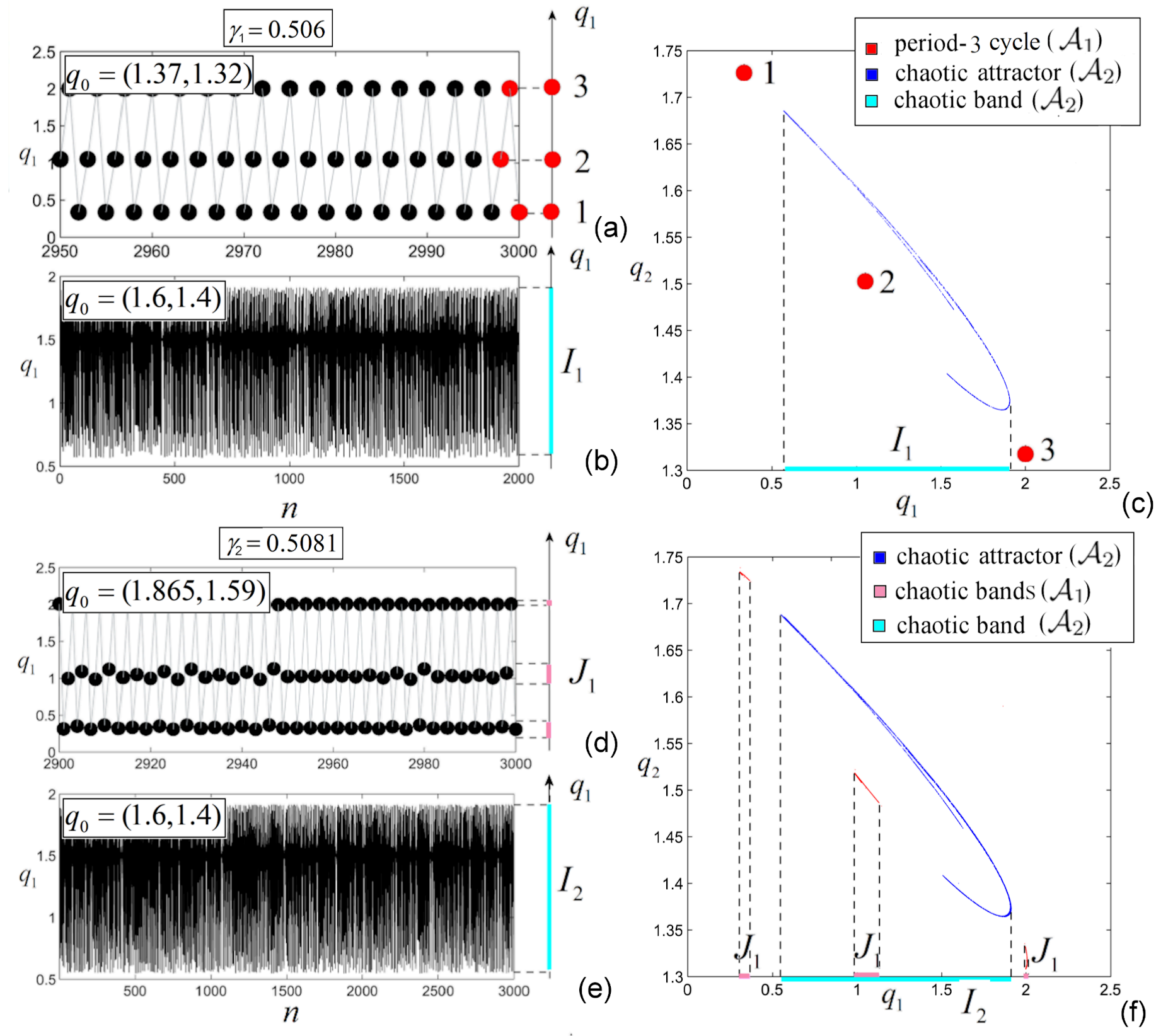}
\caption{(a) Time series of the periodic attractor $\A_1$ for $\gamma_1=0.506$ with initial condition $q_0=(1.37,1.32)$; (b) Time series of the chaotic attractor $\A_2$ for $\gamma_1=0.506$ with initial condition $q_0=(1.6,1.4)$; (c) Phase plot of attractors $\A_1$ and $\A_2$ for $\gamma=\gamma_1$; (d) Time series of the chaotic attractor $\A_1$ for $\gamma_2=0.5081$ with initial condition $q_0=(1.865,1.59)$; (e) Time series for the chaotic attractor $A_2$ for $\gamma_2=0.5081$ with initial condition $q_0=(1.6,1.4)$; (f) Phase plot of attractors $\A_1$ and $A_2$ for $\gamma=\gamma_2$.}
%\label{fig5}
\end{center}
\end{figure}

\begin{figure}
\ContinuedFloat
\begin{center}
\includegraphics[scale=0.5]{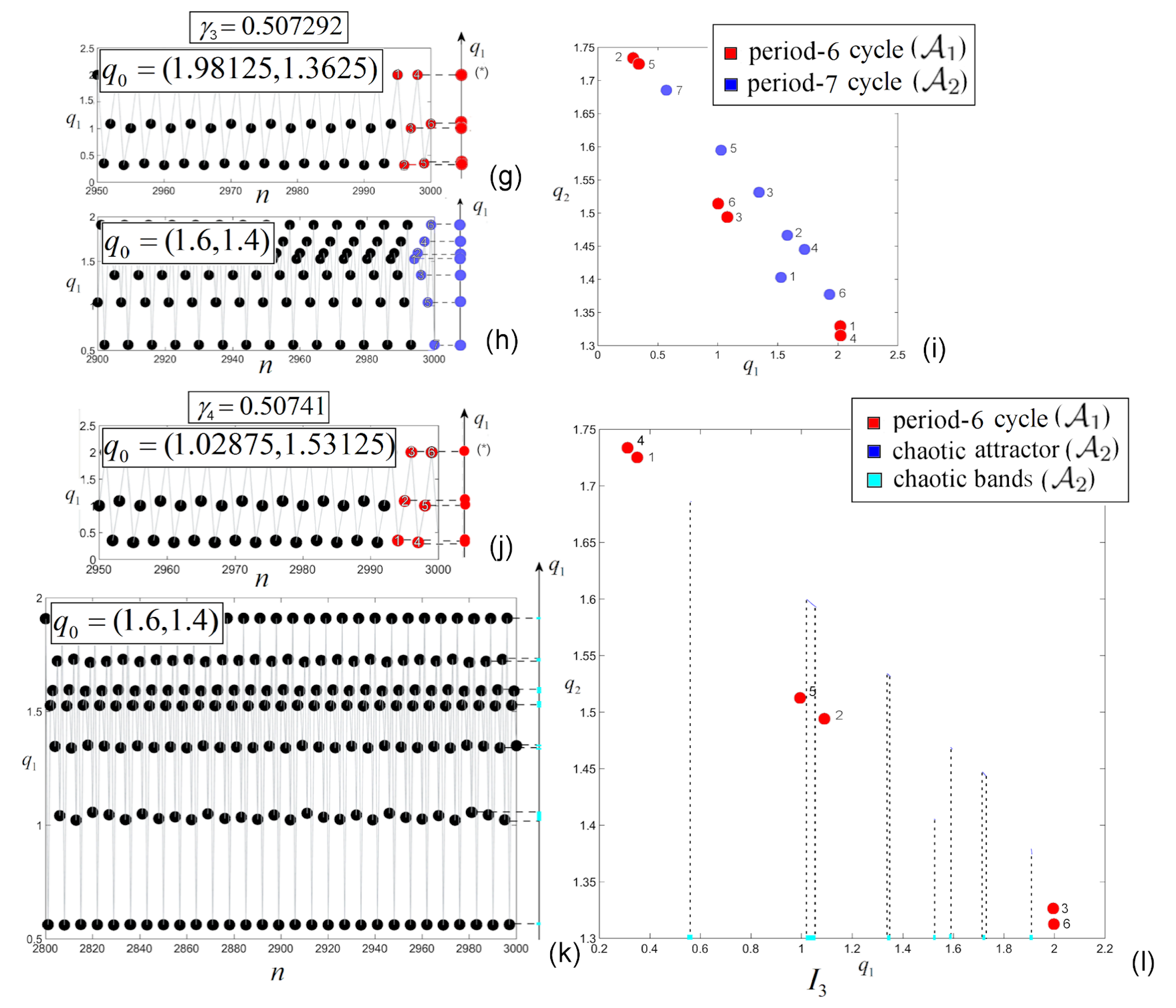}
\caption{Continuation: (g) Time series of the periodic attractor $\A_1$ for $\gamma_3=0.507292$ with initial condition $q_0=(1.98125,1.3625)$; (h) Time series of the periodic attractor $A_2$ for $\gamma_3=0.507292$; (i) Phase plot of attractors $\A_1$ and $A_2$ for $\gamma=\gamma_3$; (j) Time series of the periodic attractor $A_1$ for $\gamma_4=0.50741$ with initial condition $q_0=(1.02875,1.53125)$; (k) Time series of the chaotic attractor $A_2$ for $\gamma_4=0.50741$; (c) Phase plot of attractors $\A_1$ and $\A_2$ for $\gamma=\gamma_4$.}
\label{fig5}
\end{center}
\end{figure}

\begin{figure}
\begin{center}
\includegraphics[scale=1.3]{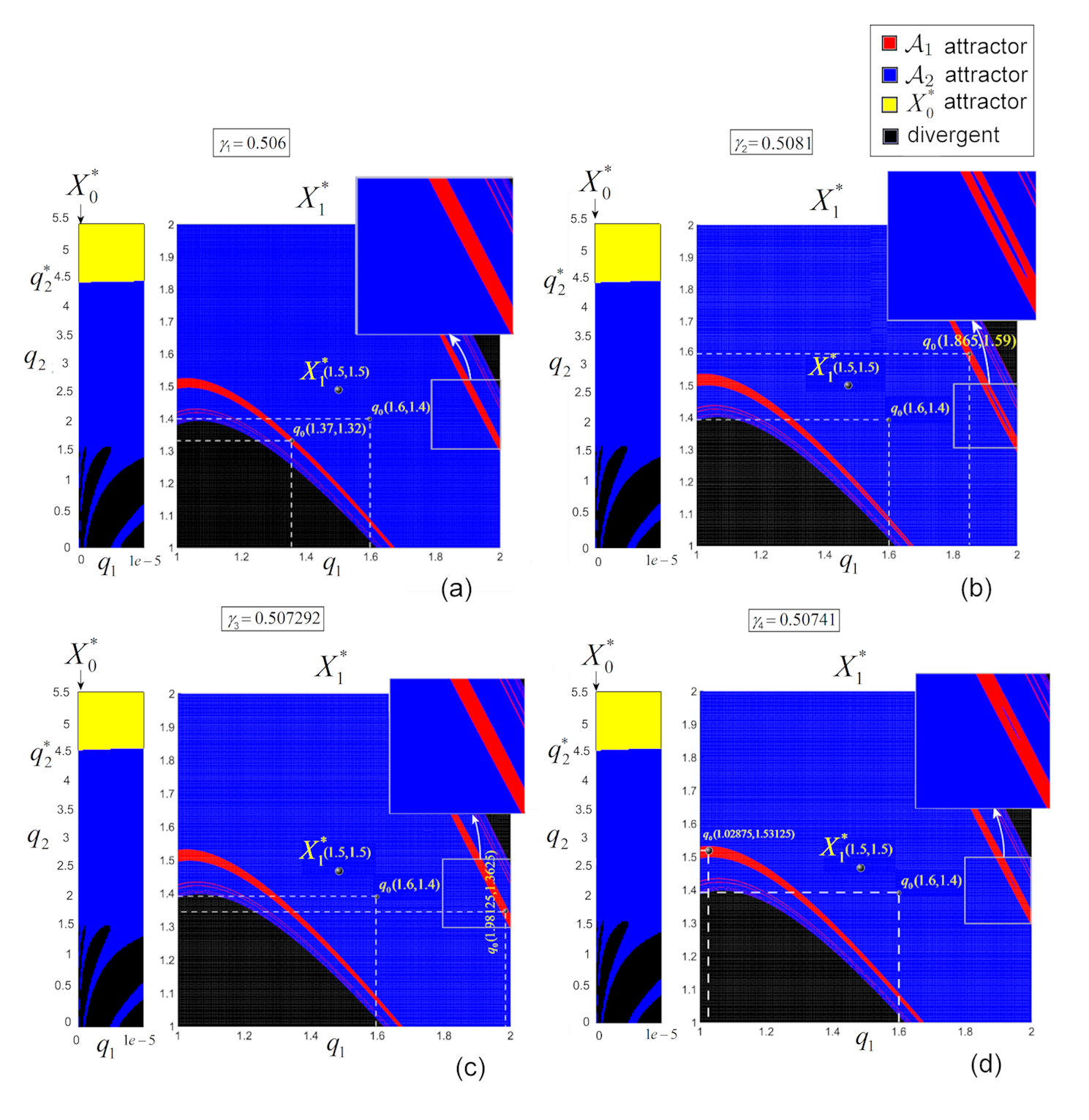}
\caption{Attraction basins of attractors $\A_1$ and $\A_2$ for the considered four $\gamma$ values, considered around both equilibria $X_{0,1}^*$. Red area represents initial conditions of the attractor $\A_1$, while blue parts of the attraction basin of $\A_2$. Yellow points are attracted by $X_0^*$ which, for $q_2>q_2^*$ is attractive. Points from black area tend to infinity. (a) $\gamma=\gamma_1$; (b) $\gamma=\gamma_2$; (c) $\gamma=\gamma_3$; (d) $\gamma=\gamma_4$.}
\label{fig6}
\end{center}
\end{figure}

\begin{figure}
\begin{center}
\includegraphics[scale=1.2]{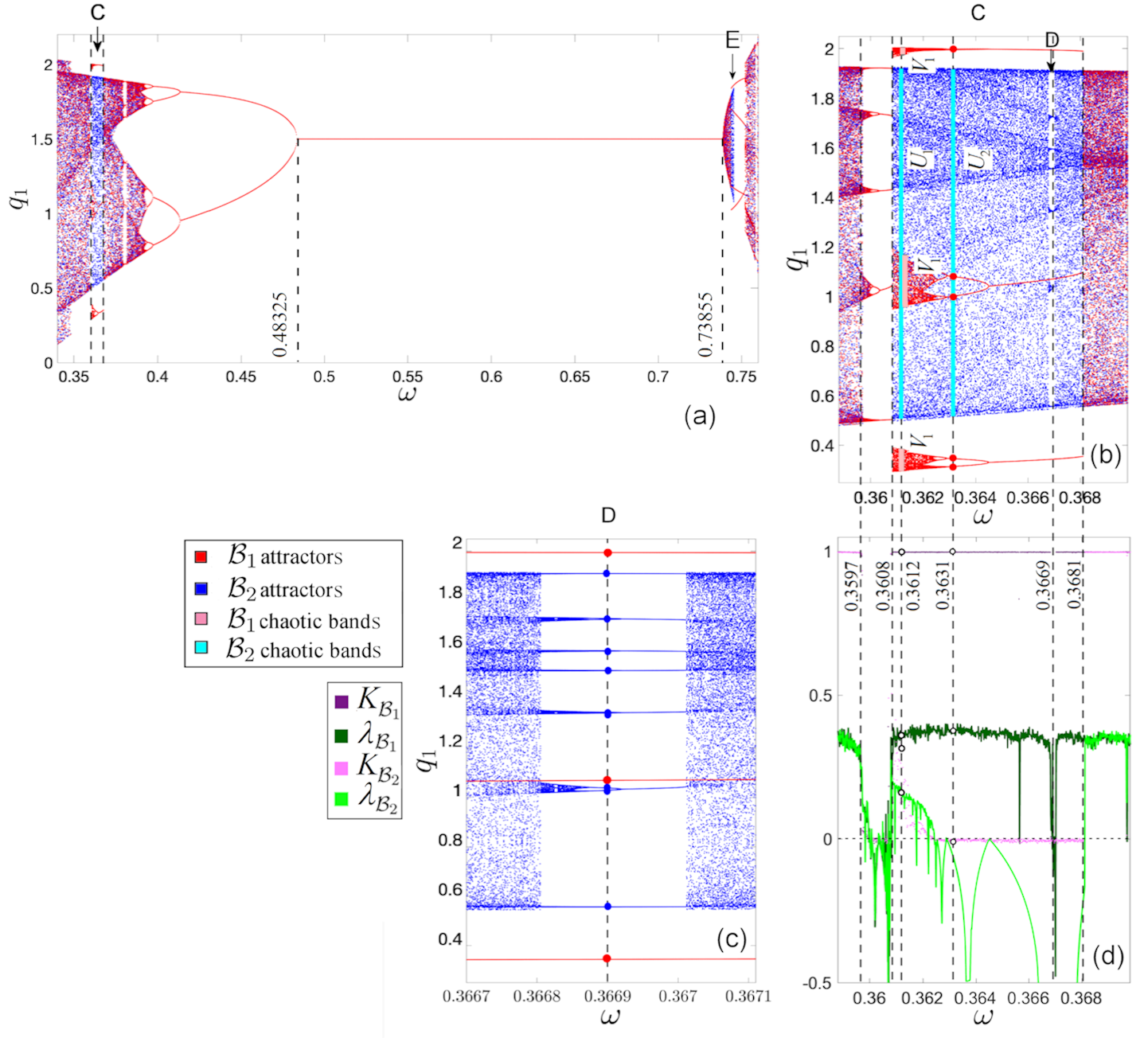}
\caption{(a) Bifurcation diagram of $q_1$ versus $\omega$, $BD_\omega$ for $\gamma=0.48$; (b) Zoomed area $C$ of the $BD_\omega$ for $\omega\in[0.3608,0.3681]$; (c) Zoomed area of $C$; (d) $K$ and $\lambda$ of the area $C$.}
\label{fig7}
\end{center}
\end{figure}

\begin{figure}
\begin{center}
\includegraphics[scale=0.42]{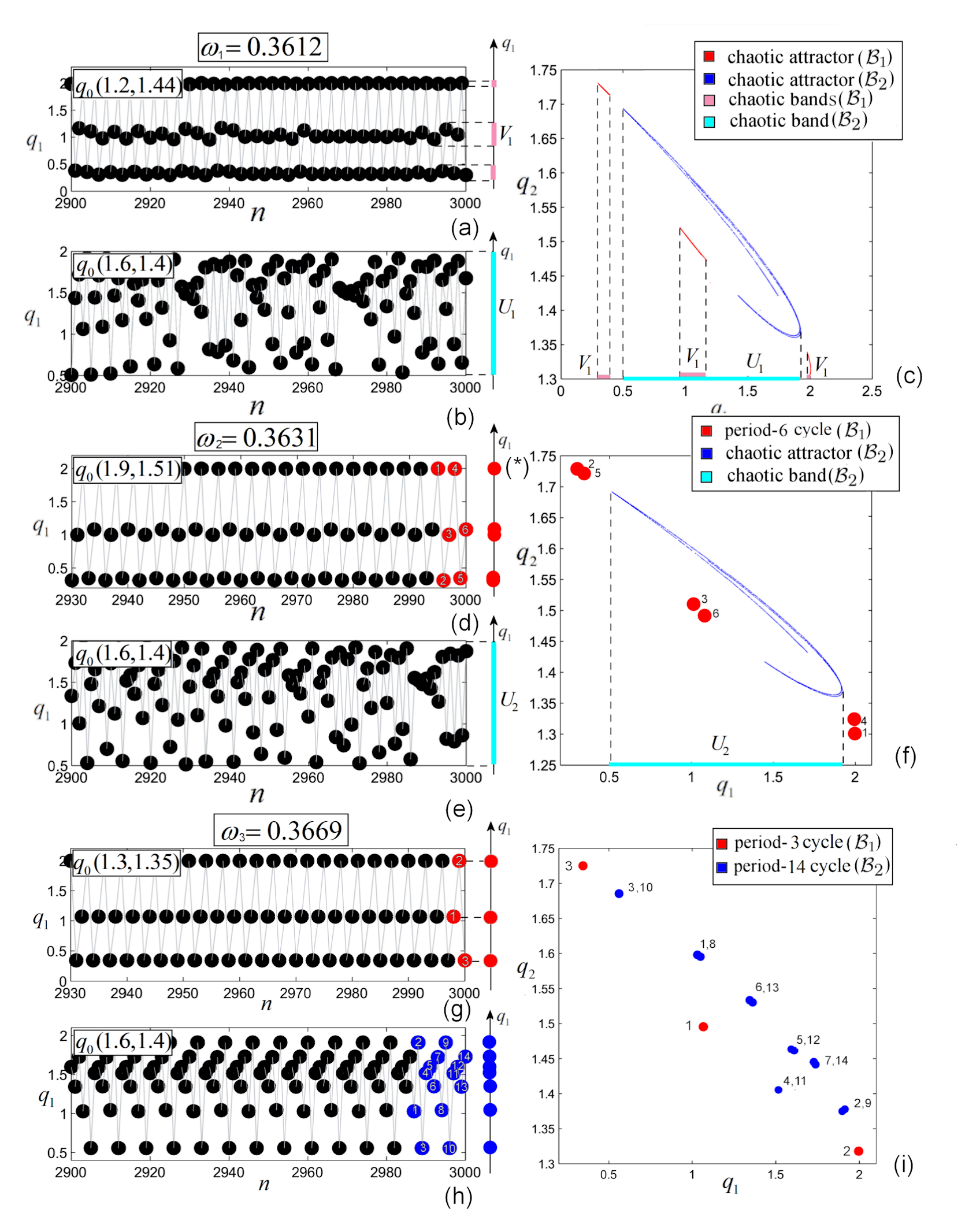}
\caption{(a) Time series for the chaotic attractor $\B_1$ for $\omega_1=0.3612$ with initial condition $q_0=(1.2,1.44)$; (b) Time series of the chaotic attractor $\B_2$ for $\omega_1=0.3612$; (c) Phase plot of attractors $\B_1$ and $B_2$ for $\omega=\omega_1$; (d) Time series of the periodic attractor $\B_1$ for $\omega_2=0.3631$ with initial condition $q_0=(1.9,1.51)$; (e) Time series of the chaotic attractor $B_2$ for $\omega_2=0.3631$; (f) Phase plot of attractors $\B_1$ and $\B_2$ for $\omega=\omega_2$; (g) Time series of the periodic attractor $\B_1$ for $\omega_3=0.3669$ with initial condition $q_0=(1.3,1.35)$; (h) Time series of the chaotic attractor $\B_2$ for $\omega=\omega_3$ with initial condition $q_0=(1.6,1.4)$; (i) Phase plot of attractors $\B_1$ and $B_2$ for $\omega=\omega_3$.}
\label{fig8}
\end{center}
\end{figure}

\begin{figure}
\begin{center}
\includegraphics[scale=0.35]{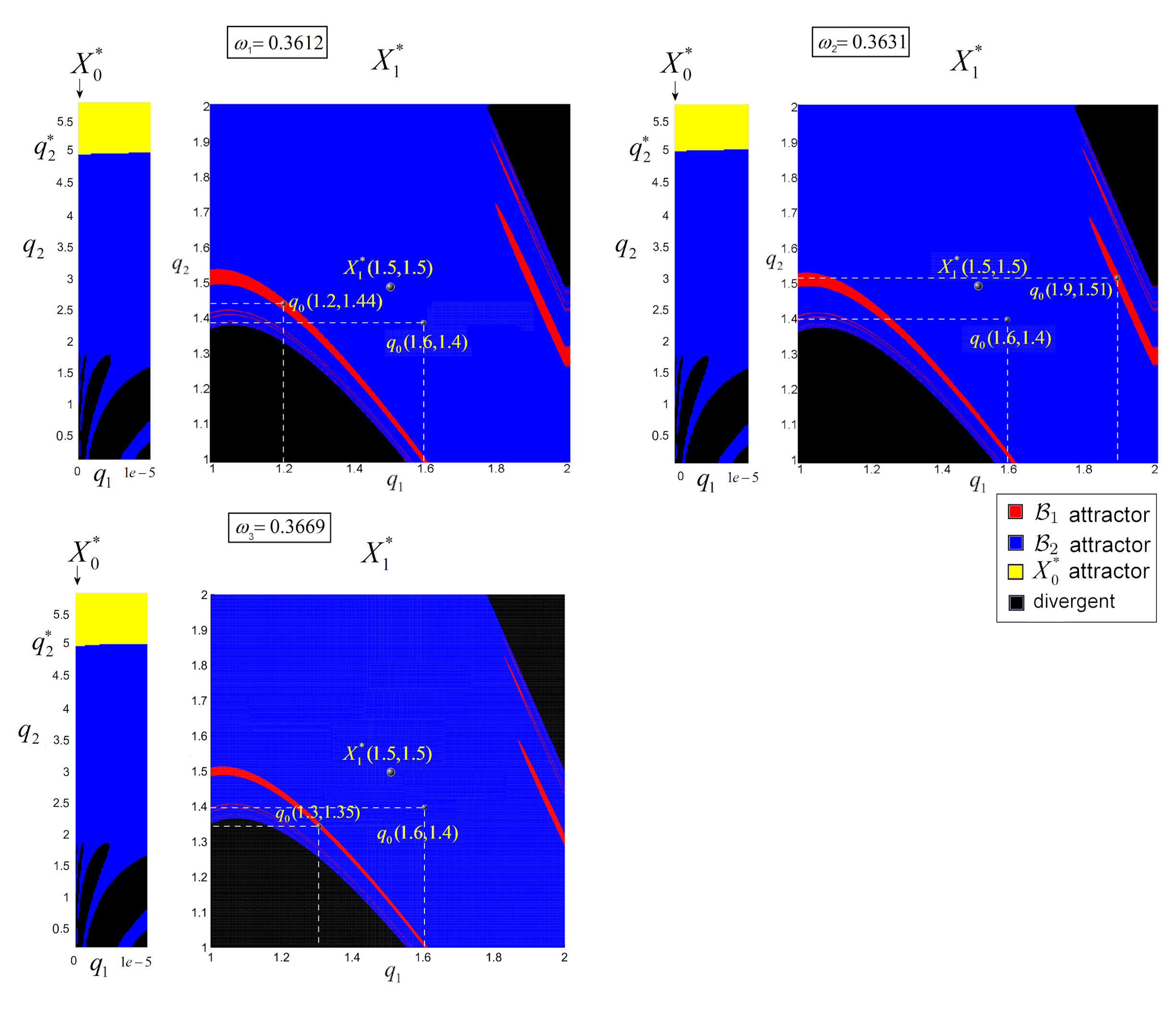}
\caption{Attraction basins of attractors $\B_1$ and $\B_2$ for the considered three $\omega$ values, considered around both equilibria $X_{0,1}^*$. Red area represents parts of the attraction basin of the attractor $\B_1$, while blue parts of the attraction basin of $\B_2$. Yellow points are attracted by $X_0^*$ which, for $q_2>q_2^*$ is attractive. Points from black area tend to infinity. (a) $\omega_1=\omega_1$; (b) $\omega=\omega_2$; (c) $\omega=\omega_3$.}
\label{fig9}
\end{center}
\end{figure}

\begin{figure}
\begin{center}
\includegraphics[scale=0.55]{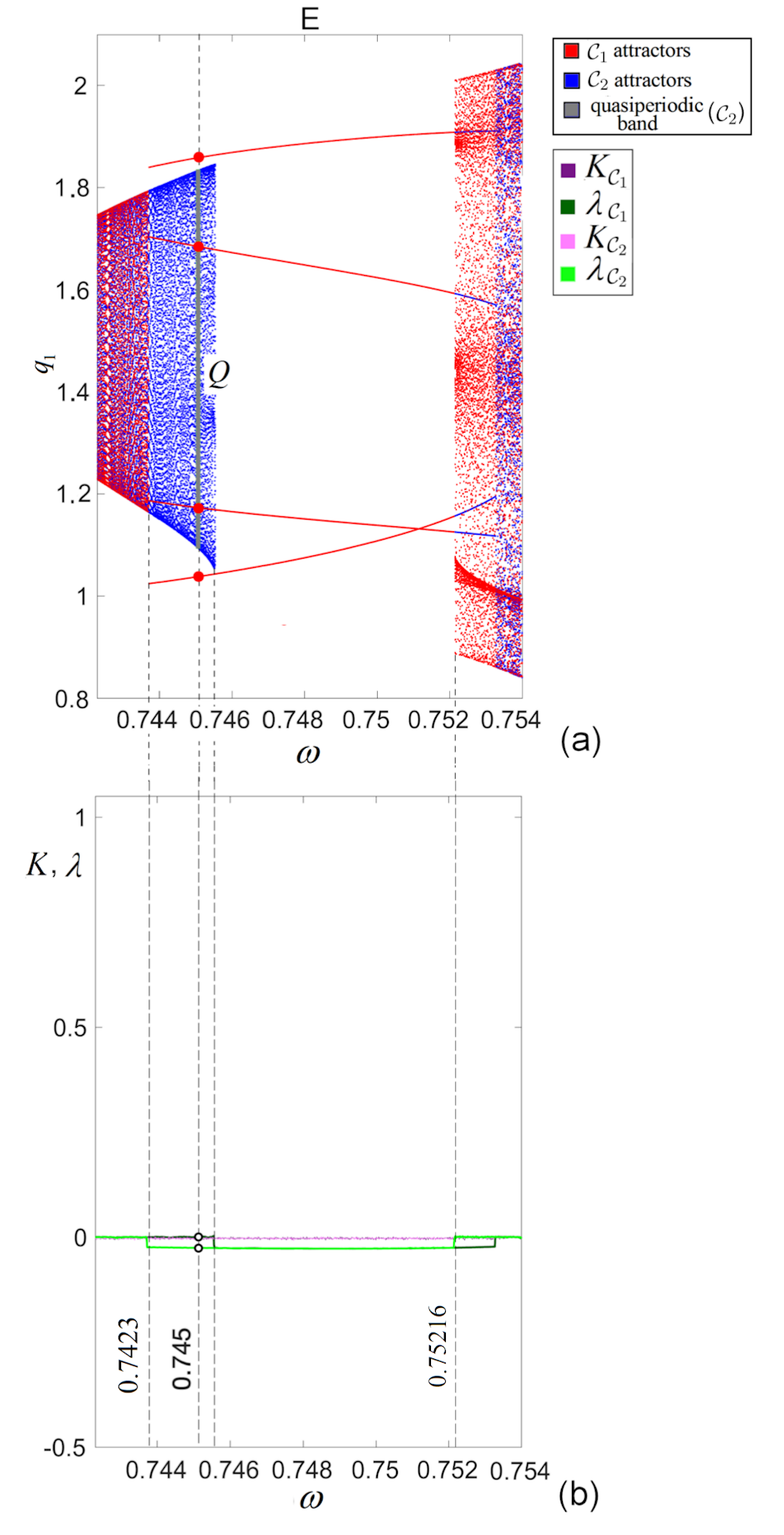}
\caption{(a) Zoomed area $E$ of the $BD_\omega$ for $\omega\in [0.7423,0.75216]$; (b) $K$ and $\lambda$.}
\label{fig10}
\end{center}
\end{figure}

\begin{figure}
\begin{center}
\includegraphics[scale=0.6]{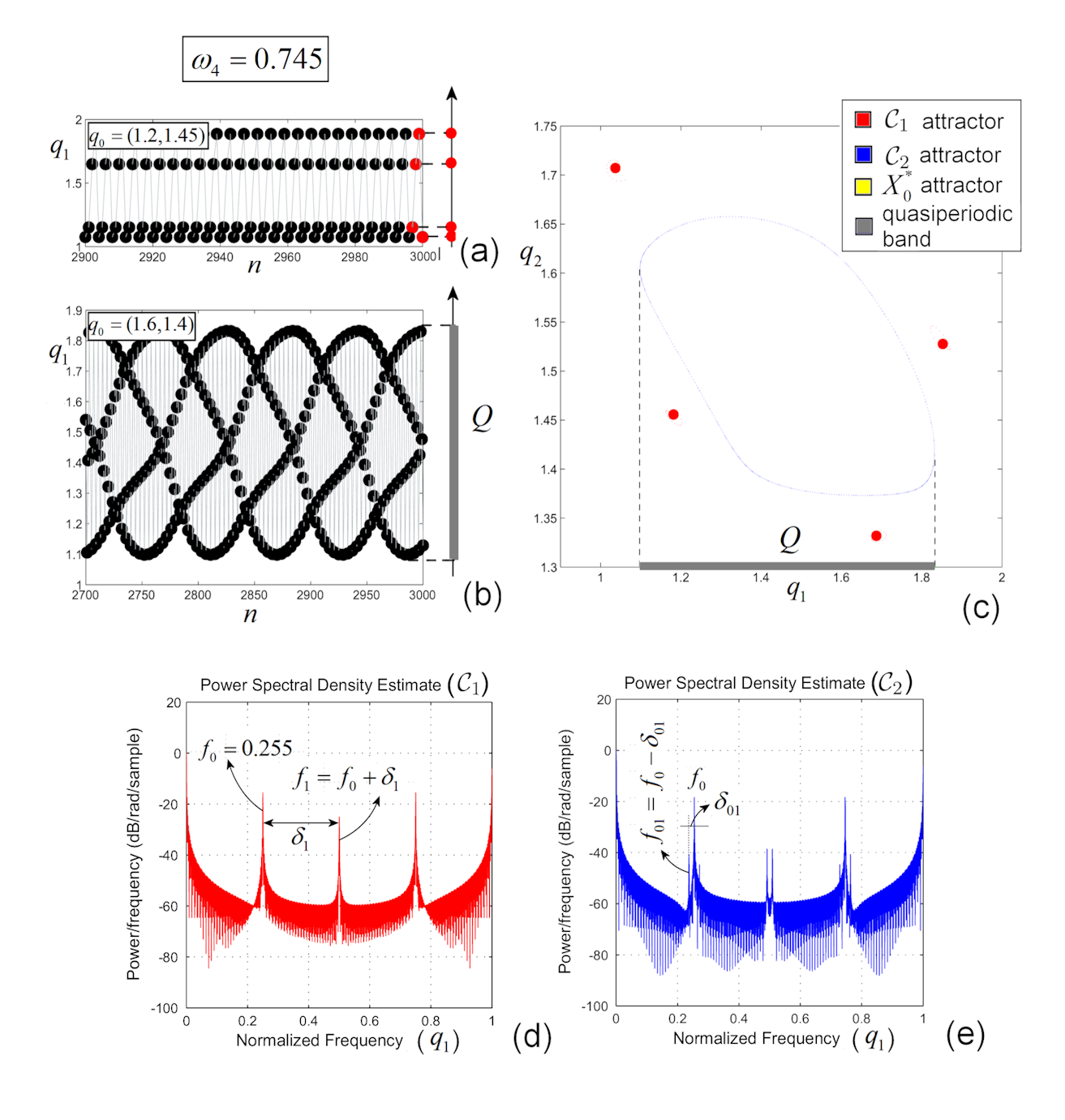}
\caption{(a) Time series of the periodic attractor $\C_1$ for $\omega_4=0.745$; (b) Time series of the quasiperiodic attractor $\C_2$; (c) Phase plot of the attractors $\C_1$ and $\C_2$; (d) PSD of the component $q_1$ of the attractor $C_1$; (e) PSD of the component $q_1$ of the attractor $C_2$. }
\label{fig11}
\end{center}
\end{figure}

\begin{figure}
\begin{center}
\includegraphics[scale=0.65]{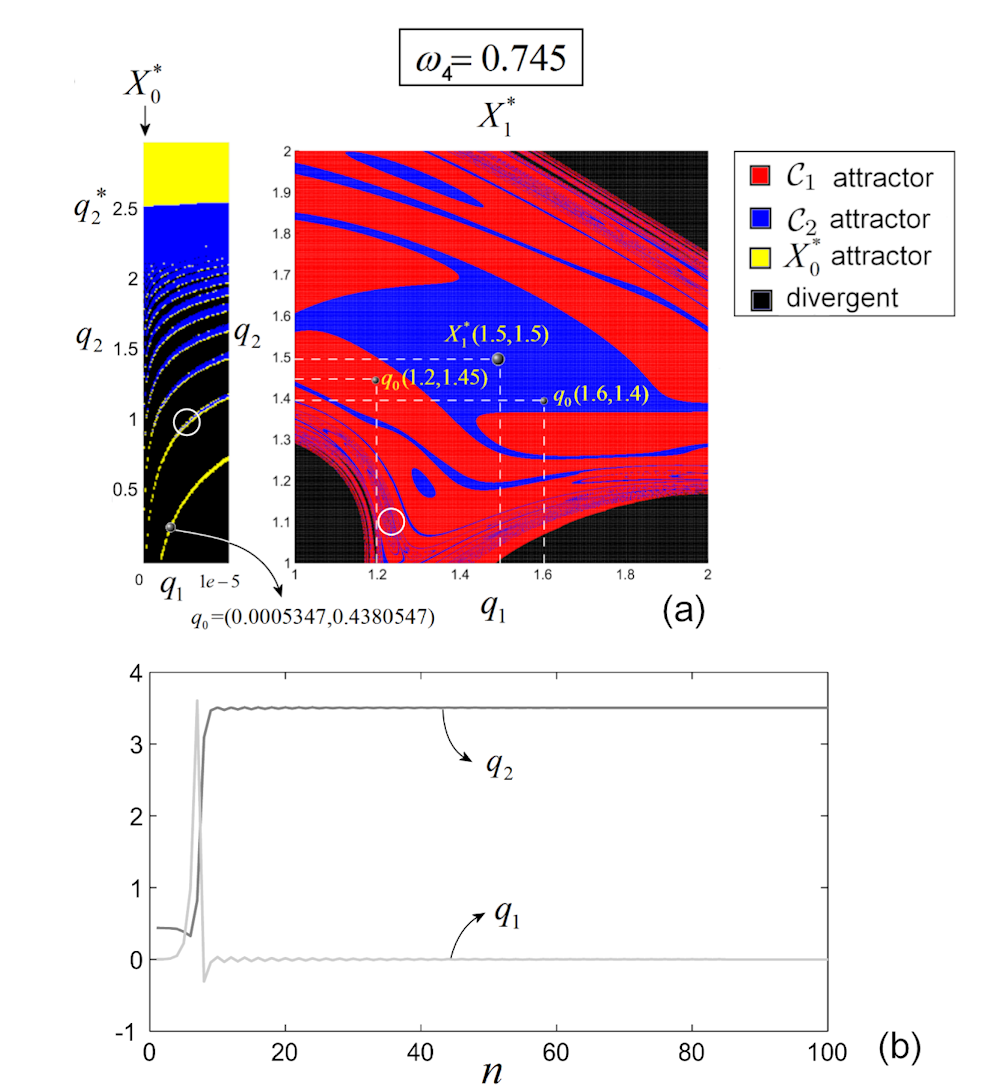}
\caption{(a) Attraction basin of attractors $\C_1$ and $\C_2$ for $\omega_4=0.745$, considered around both equilibria $X_{0,1}^*$. Red area represents parts of the attraction basin of the attractor $\C_1$, while blue parts of the attraction basin of $\C_2$. Yellow points are attracted by $X_0^*$ which, for $q_2>q_2^*$ is attractive. Points from black area tend to infinity. Circled regions reveal the complicated shape of attraction basins; (b) Orbit from $q_0=(0.0005347, 0.4380547)$ is attracted by the equilibria line $X_0^*$, even $q_0$ is within unstable area, as defined by Theorem \ref{tt} (for clarity only the first 100 iterations are shown).}
\label{fig12}
\end{center}
\end{figure}

\clearpage
%\newpage
%\bibliographystyle{plain}
%\bibliography{mybibfile}

%\bibliography{mybibfile2}

\end{document}